\documentclass[conference]{IEEEtran}
\IEEEoverridecommandlockouts

\usepackage{cite}
\usepackage{amsmath,amssymb,amsfonts}
\usepackage[T1]{fontenc}
\usepackage{graphicx}
\usepackage{textcomp}
\usepackage{xcolor}
\usepackage{enumitem}
\usepackage{flushend}
\usepackage{booktabs}
\usepackage{multirow}
\usepackage{tabularx}
\usepackage{diagbox}
\usepackage{colortbl}
\usepackage{soul}
\usepackage{wrapfig}
\usepackage{listings}
\usepackage{subcaption}
\usepackage[font=footnotesize,labelfont=bf]{caption}
\usepackage{dblfloatfix}
\usepackage{algorithm}
\usepackage[noend]{algpseudocode}
\usepackage{algcompatible}

\usepackage{bbding}
\usepackage{fontawesome5}
\newcommand{\cmark}{\textcolor{green!70!black}{\faCheckCircle}} 
\newcommand{\xmark}{\textcolor{red!70!black}{\faTimesCircle}} 
\newcommand{\warn}{\textcolor{yellow!60!red}{\faExclamationCircle}}

\def\BibTeX{{\rm B\kern-.05em{\sc i\kern-.025em b}\kern-.08em
    T\kern-.1667em\lower.7ex\hbox{E}\kern-.125emX}}

\usepackage{comment}
\usepackage[font=footnotesize,labelfont=bf]{caption}
\usepackage{tablefootnote}
\usepackage{makecell}
\usepackage{pifont}
\usepackage{flushend}
\usepackage{booktabs}
\usepackage{cite}
\usepackage{amsmath,amssymb,amsfonts,bm}
\usepackage{algorithm}

\usepackage{graphicx}
\usepackage{textcomp}
\usepackage[colorlinks=true, urlcolor=black, linkcolor=black, citecolor=black, pdfborder={0 0 0}]{hyperref}

\definecolor{lightgreen}{HTML}{DFF0D8}
\definecolor{lightred}{HTML}{F2DEDE}
\definecolor{lightyellow}{HTML}{FCF8E3}

\usepackage{algpseudocode}

\begin{document}

\title{CircuitGuard: Mitigating LLM Memorization in RTL Code Generation Against IP Leakage}

\author{\IEEEauthorblockN{Nowfel Mashnoor, Mohammad Akyash, Hadi Kamali, Kimia Azar}
\IEEEauthorblockA{\textit{Department of Electrical and Computer Engineering (ECE), University of Central Florida, Orlando, FL 32816, USA} \\
\{nowfel.mashnoor, mohammad.akyash, kamali, azar\}@ucf.edu}
}

\maketitle

\begin{abstract}

Large Language Models (LLMs) have achieved remarkable success in generative tasks, including register-transfer level (RTL) hardware synthesis. However, their tendency to memorize training data poses critical risks when proprietary or security-sensitive designs are unintentionally exposed during inference. While prior work has examined memorization in natural language, RTL introduces unique challenges: In RTL, structurally different implementations (e.g., behavioral vs. gate-level descriptions) can realize the same hardware, leading to intellectual property (IP) leakage (full or partial) even without verbatim overlap. Conversely, even small syntactic variations (e.g., operator precedence or blocking vs. non-blocking assignments) can drastically alter circuit behavior, making correctness preservation especially challenging. In this work, we systematically study memorization in RTL code generation and propose CircuitGuard, a defense strategy that balances leakage reduction with correctness preservation. CircuitGuard (i) introduces a novel RTL-aware similarity metric that captures both structural and functional equivalence beyond surface-level overlap, and (ii) develops an activation-level steering method that identifies and attenuates transformer components most responsible for memorization. Our empirical evaluation demonstrates that CircuitGuard identifies (and isolates) 275 memorization-critical features across layers 18–28 of Llama 3.1-8B model, achieving up to 80\% reduction in semantic similarity to proprietary patterns while maintaining generation quality. CircuitGuard further shows 78–85\% cross-domain transfer effectiveness, enabling robust memorization mitigation across circuit categories without retraining.
\footnote{Code is available at \href{https://github.com/mashnoor/circuitguard}{https://github.com/mashnoor/circuitguard}}.

\end{abstract}


\section{Introduction}

Large Language Models (LLMs) have demonstrated impressive proficiency across a wide range of generative tasks, from open-domain dialogue \cite{yi2024survey, deng2023prompting} to domain-specific code synthesis \cite{liu2024exploring, fakhoury2024llm}. This capability, however, comes with an inherent risk: LLMs can memorize and reproduce portions of their training data during inference, especially when prompted in ways that trigger verbatim or near-verbatim recall \cite{sakarvadia2025mitigatin, satvaty2025undesirable, biderman2023emergent, karamolegkou2023copyright}. While such memorization is a recognized concern in natural language generation, the risk becomes even more severe in the context of RTL generation, where training data may include proprietary or confidential IPs \cite{wang2025verileaky, wang2025salad}. 

Recent work demonstrates that adversaries can extract gigabytes of training data from language models across the openness spectrum, including open-source systems such as Pythia \cite{biderman2023pythia} and semi-open models such as LLaMA \cite{touvron2023llama}, and even proprietary platforms like ChatGPT \cite{nasr2023scalable}. This phenomenon stems from the fact that LLMs often memorize training data by encoding it in their parameters, enabling inference-time queries to elicit verbatim or near-verbatim reproduction \cite{sakarvadia2025mitigatin}. Such behavior poses significant IP confidentiality issues, particularly when sensitive or proprietary data is embedded in the training corpus \cite{carlini2021extracting}. To address these risks, research on memory mitigation has gained traction, aiming to prevent unintended leakage without substantially degrading model utility \cite{staab2024beyond}. Approaches range from data-centric interventions (e.g., dataset deduplication \cite{carlini2022quantifying}) to training-level defenses (e.g., differential privacy \cite{singh2024whispered}). A promising complementary direction is machine unlearning, which seeks to selectively remove or suppress the influence of specific data samples from an already-trained model, thereby offering a targeted remedy against memorization-driven leakage \cite{liu2025rethinking, usynin2024memorisation}.

\begin{table}[t]
\centering
\setlength{\tabcolsep}{0.2pt}
\caption{Comparison of RTL Leakage Mitigation Approaches.}
\label{tab:top_comparison}
\begin{tabular}{@{} l *{4}c @{}}
\toprule
\textbf{Dimension}  & \textbf{VeriLeaky \cite{wang2025verileaky}} & \textbf{SALAD \cite{wang2025salad}} & \textbf{CircuitGuard (prop.)} \\
\cmidrule(r){1-1} \cmidrule(r){2-2} \cmidrule(r){3-3} \cmidrule(r){4-4}
\multirow{2}{*}{\textbf{Method}} & Similarity Check & Machine &  Unlearning w/\\
 & \& Obfuscation & Unlearning & Activation steering \\
\cmidrule(r){1-1} \cmidrule(r){2-2} \cmidrule(r){3-3} \cmidrule(r){4-4}
\textbf{Time of} & Before & Training & Training \& Inference \\ 
\textbf{Applying} & Fine-tuning & Time + Retraining & Time \\
\cmidrule(r){1-1} \cmidrule(r){2-2} \cmidrule(r){3-3} \cmidrule(r){4-4}
\textbf{Stops} & \xmark~Leaking Design & \warn~Subset & \cmark~Suppresses\\
\textbf{Leakage?} &  \& Obfuscation & Removal (not Full) & Memorization \\ 
\cmidrule(r){1-1} \cmidrule(r){2-2} \cmidrule(r){3-3} \cmidrule(r){4-4}
\textbf{Correctness} & \warn~Degraded by & \cmark~Affected w/ & \multirow{2}{*}{\cmark~ Tunable} \\ 
\textbf{Guarantee} & Obfuscation & Over-forgetting\\
\cmidrule(r){1-1} \cmidrule(r){2-2} \cmidrule(r){3-3} \cmidrule(r){4-4}
\textbf{Scalability} & \cmark~No Retraining & \warn~Needs Retraining & \cmark~Minimal Steering \\
\cmidrule(r){1-1} \cmidrule(r){2-2} \cmidrule(r){3-3} \cmidrule(r){4-4}
\textbf{Ease of} & \xmark~Needs Obfuscation & \warn~Costly Retrain & \cmark~Plug-in \\
\textbf{Deployment} & Implementation & w/ Unlearning & Steering \\
\bottomrule
\end{tabular}
\end{table}


Despite growing awareness of memorization risks in LLMs, the problem remains largely unexplored for hardware design. In RTL, memorization can directly lead to IP leakages of critical blocks, e.g., cryptographic cores \cite{chhotaray2017standardizing}. Unlike natural language, where memorization raises copyright or privacy issues \cite{meeus2024copyright, eldan2023whos}, leaked RTL can undermine security guarantees and compromise years of engineering investment \cite{pilato2021assure, kamali2022advances}. Recent work such as VeriLeaky \cite{wang2025verileaky} highlights how fine-tuning on proprietary IP leads to measurable leakage of sensitive RTL modules, even under obfuscation/locking techniques. Also, SALAD \cite{wang2025salad} shows that machine unlearning can selectively remove contaminated, proprietary, or malicious RTL subsets from LLMs, restoring trust while preserving generalization. However, they rely on lexical or semantic overlap, which fail to capture the complexity of RTL similarity \cite{akyash2025simeval}. Functionally equivalent designs may look very different due to naming conventions or synthesis-driven transformations, yet still leak IP. These limitations underscore the need for unlearning-based defenses tailored to RTL that go beyond surface-level overlap and actively \textit{forget} sensitive subsets, while ensuring that mitigation does not deteriorate the correctness of generated RTL.

In RTL generation, three key challenges of memorization must be addressed:
\textbf{\textit{\underline{(i) Selective Forgetting of RTL Subsets:}}} Training LLMs on RTL corpora risks unintended reproduction of proprietary or sensitive designs. Even when structural transformations (e.g., refactoring, renaming, or logic obfuscation) disguise overlap, latent similarities still lead to memorization and leakage.
\textbf{\textit{\underline{(ii) Defining RTL-Aware Forgetting Metrics:}}} Effective unlearning requires principled similarity metrics that go beyond surface-level token overlap to capture structural and functional equivalence, enabling precise detection of whether sensitive designs have been “forgotten.”
\textbf{\textit{\underline{(iii) Preserving Generalization and Correctness:}}} Forgetting one subset must not degrade the model’s ability to generate correct and reliable RTL on the remaining (non-sensitive) data.

To do so, we contribute the following:

\noindent \textbf{(1) Attention-steered unlearning:} \textbf{CircuitGuard} identifies attention components within transformer layers most responsible for memorization of the sensitive subset and develops an inference-time steering mechanism to attenuate their influence, thereby selectively “unlearning” proprietary patterns.

\noindent \textbf{(2) RTL-aware forgetting metric:} We design a similarity metric tailored for RTL code that captures both structural and functional equivalence, allowing us to measure whether memorization of sensitive designs has been effectively reduced.   

\noindent \textbf{(3) Empirical validation of unlearning:} By the metric, we show through extensive experiments that our method successfully unlearns proprietary-marked subsets while maintaining correctness and generalization on the non-sensitive subset.  

\section{Related Works}

\begin{figure}[t]
\centering
\includegraphics[width=0.95\linewidth]{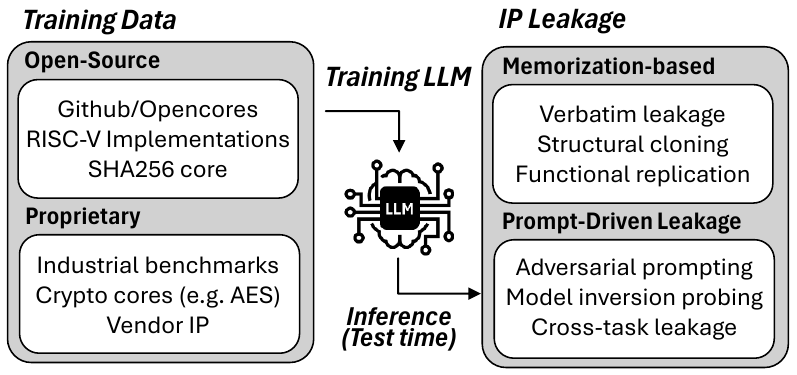}
\caption{Overview of IP Leakage Pathways in LLM-Driven RTL Generation}
\label{fig:overview_ip_leakage}
\end{figure}

Figure~\ref{fig:overview_ip_leakage} illustrates the potential for IP leakage in LLM-aided RTL design, where training on mixed open-source and proprietary datasets can result in inference-time disclosures. This section reviews prior work on LLMs for RTL generation, noting technical advances (e.g., synthesis fidelity, semantic integration) and security concerns that call for defenses against memorization-driven IP leakage.

\subsection{LLM for RTL Code Generation} 

The use of LLMs for hardware design has gained momentum in recent years, with several efforts targeting RTL code generation \cite{thakur2024verigen, liu2024rtlcoder, akyash2025rtl++, zhao2024codev}. VeriGen \cite{thakur2024verigen} introduced one of the first large-scale datasets of RTL designs, collected from open-source repositories, to enable instruction-to-code synthesis.  Building on this foundation, RTLCoder \cite{liu2024rtlcoder} demonstrated that LLMs can synthesize RTL designs directly from natural language specifications, highlighting the potential for automating hardware design entry. 
RTL++ \cite{akyash2025rtl++} further advanced the field by incorporating graph-based structural information, improving the functional guarantee of generated designs through integration of semantic and structural cues. More recently, CodeV \cite{zhao2024codev} proposed a multi-stage summarization framework that captures hierarchical RTL semantics, leading to higher quality code outputs. Despite extensive fine-tuning, risks like IP leakage remain underexplored, indicating the need for methods that curb memorization while preserving RTL correctness.

\subsection{Memorization in LLMs and Mitigation Approaches}  

\begin{figure*}[t]
\centering
\includegraphics[width=\linewidth]{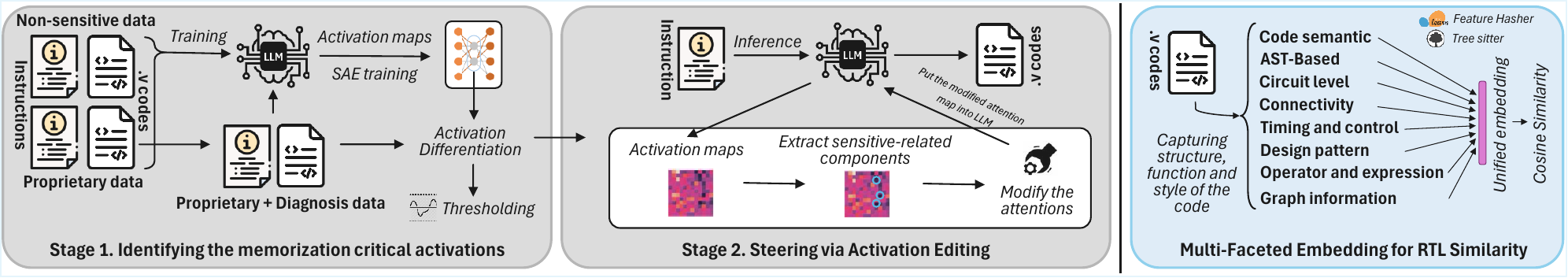}
\caption{Overview of the Proposed Framework (CircuitGuard). Stage 1 uses sparse autoencoders on proprietary and diagnostic data to identify memorization-critical activations. Stage 2 edits these activations during inference to suppress sensitive patterns in generated RTL. A multi-faceted embedding—covering semantics, AST, circuit structure, connectivity, timing/control, and design patterns—measures RTL similarity for leakage evaluation.}
\vspace{-15pt}
\label{fig:CircuitGuard_framework}
\end{figure*}

Several studies have demonstrated that LLMs can memorize and reproduce specific training samples \cite{carlini2021extracting, biderman2023emergent, nasr2023scalable}. Memorization occurs in both verbatim and paraphrased forms, with targeted prompts shown to elicit sensitive content such as personal identifiers or proprietary code \cite{sakarvadia2025mitigatin, satvaty2025undesirable}. Early work by Carlini et al.~\cite{carlini2021extracting} provided systematic evidence of extraction attacks, while subsequent studies revealed that memorization is widespread across open-source models (e.g., Pythia \cite{biderman2023pythia}) as well as proprietary platforms like ChatGPT \cite{nasr2023scalable}. The results suggest that memorization does not remain consistent throughout training; models show increased memorization during the early and final stages, with the lowest levels observed around the midpoint \cite{leybzon2024learning}. These observations underscore the dual-use nature of LLMs, where their generative capabilities can unintentionally compromise confidentiality.

Several classes of defenses have been proposed. Data-centric interventions such as dataset deduplication aim to reduce redundant memorization during pretraining \cite{carlini2022quantifying}. Training-level defenses, including differential privacy \cite{singh2024whispered}, adversarial regularization, and gradient noise, suppress parameter-level retention of sensitive data. Inference-time safeguards, such as activation steering \cite{suri2025mitigatin}, mitigate leakage by detecting or suppressing memorized continuations \cite{staab2024beyond}. More recently, machine unlearning has emerged as a complementary paradigm, enabling selective removal of specific training samples from already-trained models \cite{usynin2024memorisation, liu2025rethinking}. Although effective in natural language, these methods depend on surface-level similarity, which fails to capture the correctness and functionality required in RTL.

\textbf{Memorization in RTL:} VeriLeaky \cite{wang2025verileaky} provides the first systematic study of IP leakage in LLM-driven RTL coding. By fine-tuning LLaMA 3.1-8B with RTLCoder plus proprietary IP, the authors show that models can regenerate sensitive modules with up to 46.5\% similarity, confirmed via structural (AST/Dolos) and functional (Synopsys Formality) checks. They also explore logic locking as a defense, but find it reduces model utility. SALAD \cite{wang2025salad} complements this by applying machine unlearning to remove contaminated benchmarks, proprietary designs, or malicious payloads without full retraining. It demonstrates that preference-based and representation-level unlearning (e.g., SimNPO, RMU) balance forgetting and utility better than aggressive gradient-based methods. A high-level comparison of VeriLeaky, SALAD, and our proposed \textbf{CircuitGuard} framework is provided in Table~\ref{tab:top_comparison}, highlighting how our method enables inference-time, RTL-based selective forgetting while preserving correctness and utility.

\section{CircuitGuard: Methodology}

Figure~\ref{fig:CircuitGuard_framework} summarizes the complete CircuitGuard methodology. The framework integrates the three pillars: (i) dataset partitioning into non-sensitive, proprietary-marked, and diagnostic subsets, (ii) identification of memorization-critical activations using sparse autoencoders, and (iii) inference-time steering through targeted activation editing. Complementing these, our multi-faceted embedding provides a robust metric for detecting leakage and functional overlap between generated and reference RTL designs.

\subsection{Data Collection and Partitioning}

\begin{figure}[t]
\centering
\includegraphics[width=.75\linewidth]{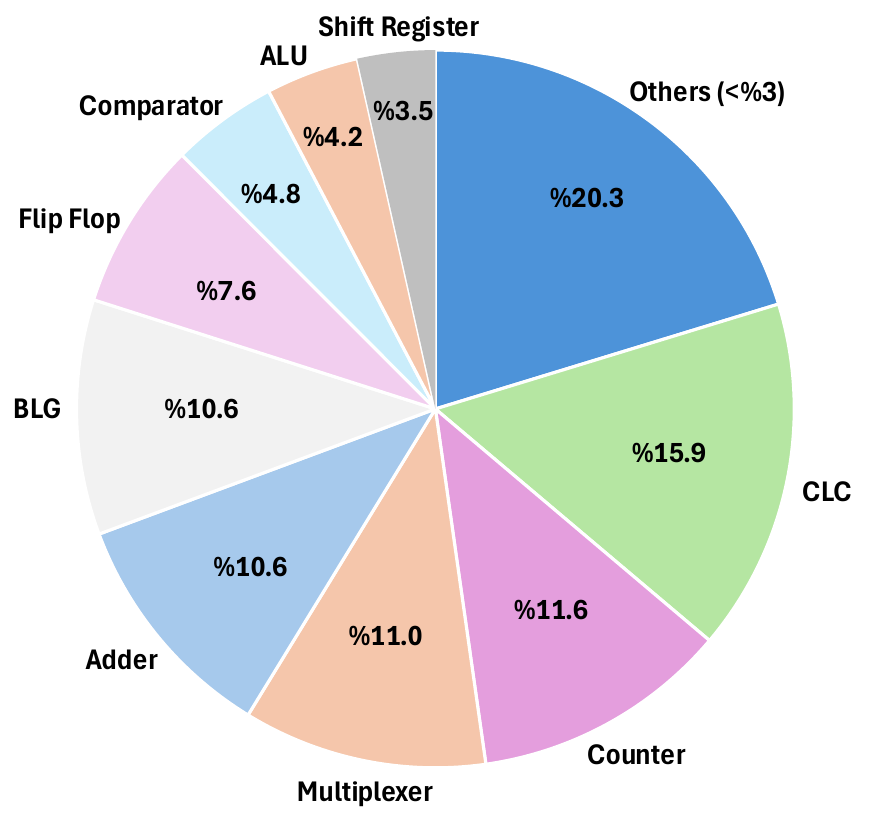}
\caption{Category Distribution of Dataset used in CircuitGuard.}
\label{fig:dataset_distribution}
0\end{figure}

To systematically study memorization and leakage risks in RTL code generation, we first construct a dataset of RTL files collected from large-scale open-source repositories (e.g., GitHub, OpenCores, and benchmark suites). The corpus spans a wide range of hardware designs, including arithmetic units, control modules, memory controllers, communication interfaces, and cryptographic primitives. The category distribution of this dataset is illustrated in Figure~\ref{fig:dataset_distribution}. Since our goal is to analyze when memorization becomes harmful, we partition the dataset into two complementary subsets (plus testing subset):

\noindent \textbf{(1) Non-Sensitive Subset:} Consists of approximately 1,700 modules that are extensively available in the public domain and whose reproduction poses minimal security or intellectual property risk. Examples include canonical benchmarks (e.g., ISCAS, ITC’99), widely reused arithmetic blocks (e.g., adders, multipliers), and simple controllers.

\noindent \textbf{(2) Proprietary-Marked Subset:} Contains around 300 modules as sensitive or proprietary, such as cryptographic cores, domain-specific accelerators, etc. While collected from open sources, we annotate these designs as ``proprietary-marked'' to emulate realistic scenarios where similar IP belongs to industrial partners or internal corporate flows. 

\noindent \textbf{(3) Hold-Out Diagnostic Subset:} A separate set of 100 modules that are never included in training. This subset is used exclusively for probing activation patterns and identifying transformer components most correlated with memorization.

\subsection{Identifying Memorization-Critical Activations}

With the dataset, we train our model on the combined set of 2{,}000 samples (1{,}700 non-sensitive and 300 proprietary-marked). To localize the internal representations most responsible for memorization, we probe the model using a set of 200 inputs: 100 proprietary-marked and 100 diagnostic.

Let $x \in \mathcal{X}$ denote an RTL entry and $h_\ell(x) \in \mathbb{R}^{d}$ the hidden activation at layer $\ell \in \{1,\dots,L\}$\footnote{E.g., \# of layers $L{=}28$, and hidden size $d{=}4096$ for LLaMA 3.1 8B.}. We first train a \emph{sparse autoencoder} (SAE) on these activations:
\[
z_\ell = f_\ell(h_\ell), 
\quad \hat{h}_\ell = g_\ell(z_\ell),
\]
with reconstruction loss
\[
\min_{\Theta_\ell} \;\; \mathbb{E}_{x}\!\left[\;\|h_\ell(x) - g_\ell(f_\ell(h_\ell(x)))\|_2^2 \right] + \lambda \|z_\ell(x)\|_1 ,
\]
where $f_\ell,g_\ell$ are encoder/decoder, $\Theta_\ell$ are SAE parameters, and $\lambda$ enforces sparsity. After training the SAE, we feed both proprietary-marked ($\mathcal{P}$) and diagnostic ($\mathcal{D}$) subsets through the model to obtain latent codes $z_\ell(x)$. For each latent dimension $i$, we compute the mean activation difference:
\[
\Delta_{\ell,i} = \left| \frac{1}{|\mathcal{P}|}\sum_{x \in \mathcal{P}} z_{\ell,i}(x) \;-\; \frac{1}{|\mathcal{D}|}\sum_{x \in \mathcal{D}} z_{\ell,i}(x) \right|.
\]

If $\Delta_{\ell,i}$ exceeds a threshold $\tau$, we label the $i$-th component at layer $\ell$ as \textit{memorization-critical}. Formally,
\[
\mathcal{M}_\ell = \{\, i \;|\; \Delta_{\ell,i} \ge \tau \,\}.
\]

The resulting sets $\{\mathcal{M}_\ell\}$ capture those latent directions that are disproportionately activated by proprietary data compared to diagnostic data. These activation components serve as the targets for intervention in our CircuitGuard steering method.

\subsection{CircuitGuard Steering via Activation Editing}

With the memorization-critical indices $\mathcal{M}_\ell$ identified, CircuitGuard intervenes at inference time to suppress their influence. The procedure is as follows: for a given prompt $x$, we extract hidden activations $h_\ell(x)$ from each layer $\ell$ and encode them with the trained SAE:
\[
z_\ell = f_\ell(h_\ell(x)).
\]

For each latent coordinate $i \in \mathcal{M}_\ell$, we apply a suppression coefficient $\alpha \in [0,1]$ on the encoded hidden activations:
\[
z'_{\ell,i} = (1-\alpha)\, z_{\ell,i}, \qquad i \in \mathcal{M}_\ell ,
\]

For all coordinates that do not belong to the suppressed set $\mathcal{M}_\ell$, we leave the hidden activations untouched. In other words, if index $i$ is not marked as critical, then the output coordinate simply remains the same as the input:
\[
z'_{\ell,i} = z_{\ell,i}, \qquad i \notin \mathcal{M}_\ell .
\]

The modified latent, after suppression has been applied, is subsequently passed through the decoder (or the next stage of the network) to map it back into the activation space.
\[
h'_\ell(x) = g_\ell(z'_\ell),
\]
The reconstructed activations replace the original $h_\ell(x)$ within the forward propagation of the language model. This intervention is applied at each chosen layer, so that the suppression is done in a layer-wise manner throughout the transformer.

By selectively attenuating only the memorization-critical components, CircuitGuard reduces the probability of proprietary IP regurgitation without disrupting the majority of latent directions. As a result, the model preserves functional correctness and generative diversity on non-sensitive designs while mitigating leakage on proprietary-marked ones.

\subsection{Multi-Faceted Embedding for RTL Similarity}
\begin{algorithm}[t]
\caption{RTL Code Embedding Generation with Feature Weights}
\label{alg:verilog_similarity}
{\small\begin{algorithmic}
\REQUIRE Query code $q$, Corpus embeddings $\mathbf{E} \in \mathbb{R}^{n \times d}$, $k$
\ENSURE Top-$k$ similar codes with scores

\STATE \textcolor{red!80!black}{\textbf{// Define features with weights}}
\STATE $\mathcal{F} \gets \{(\bm{e}_{sem}, w_{sem}), (\bm{f}_{ast}, w_{ast}), (\bm{f}_{circ}, w_{circ}),$ \\
\hspace{1.9cm} $(\bm{f}_{conn}, w_{conn}), (\bm{f}_{time}, w_{time}), (\bm{f}_{pat}, w_{pat}),$ \\
\hspace{1.9cm} $(\bm{f}_{op}, w_{op}), (\bm{f}_{lex}, w_{lex}), (\bm{f}_{graph}, w_{graph})\}$

\STATE \textcolor{red!80!black}{\textbf{// Process each feature}}
\FOR{each $(\bm{f}_i, w_i) \in \mathcal{F}$}
    \IF{$\bm{f}_i$ is sparse}
        \STATE $\bm{h}_i \gets \text{FeatureHash}(\bm{f}_i, d_i)$
    \ELSE
        \STATE $\bm{h}_i \gets \bm{f}_i$ \COMMENT{Already dense (e.g., semantic embedding)}
    \ENDIF
    \STATE $\bm{h}_i \gets \bm{h}_i / \|\bm{h}_i\|_2$ \COMMENT{L2 normalize}
    \STATE $\bm{h}_i \gets w_i \cdot \bm{h}_i$ \COMMENT{Apply feature weight}
\ENDFOR

\STATE \textcolor{red!80!black}{\textbf{// Combine weighted features}}
\STATE $\bm{v}_q \gets \bigoplus_i \bm{h}_i$ \COMMENT{Concatenate all weighted embeddings}

\STATE  \textcolor{red!80!black}{\textbf{// Compute similarities and rank}}
\STATE $\bm{s} \gets \mathbf{E} \cdot \bm{v}_q^T / (\|\mathbf{E}\|_2 \cdot \|\bm{v}_q\|_2)$ \COMMENT{Cosine similarity}
\STATE $\text{indices} \gets \text{top-k}(\bm{s})$
\STATE \textbf{return} $\{(\text{corpus}[i], \bm{s}[i]) : i \in \text{indices}\}$

\end{algorithmic}}
\end{algorithm}

To evaluate leakage and functional overlap between generated and reference RTLs, we introduce a multi-faceted embedding algorithm (see Algorithm \ref{alg:verilog_similarity}) that captures structural, behavioral, and stylistic properties of hardware designs. Unlike single-view similarity measures, our approach integrates both semantic and structural signals into \textbf{\textit{a unified vector representation}}. The embedding is built from the following components:

\noindent \textit{\underline{\textbf{(1) Semantic Code Embedding:}}} The raw RTL codes are encoded by a pre-trained sentence-transformer~\cite{reimers2019sentencebert}. This produces a dense semantic embedding that captures RTL's high-level functionality, textual semantics, and stylistic cues.

\noindent \textit{\underline{\textbf{(2) AST-Based Features:}}} Using \texttt{tree-sitter} \cite{tree-sitter}, we extract syntactic features from the abstract syntax tree (AST), including node types, $n$-grams of production rules, propagation/hierarchy, and edge counts. It captures fine-grained syntactic patterns that go beyond surface-level token similarity.

\noindent \textit{\underline{\textbf{(3) Circuit-Level Features:}}} We encode structural aspects of the circuit, such as the number of \texttt{always-ff} blocks, module count, and instantiations. These features reflect hardware hierarchy and design modularity. 

\noindent \textit{\underline{\textbf{(4) Connectivity Features:}}} To capture interface-level characteristics, we extract statistics such as input count, output count, I/O ratio, bidirectional port count, and total number of ports. These properties (particularly at modules' edges) quantify the external connectivity and I/O complexity of the design.

\noindent \textit{\underline{\textbf{(5) Timing and Control Features:}}} We analyze sequential and control constructs, including pipeline stages, register stages, delay assignments, and control-flow keywords (\texttt{if}, \texttt{else}, \texttt{for}). These features capture timing behaviors.

\noindent \textit{\underline{\textbf{(6) Design Pattern Features:}}} We detect recurring RTL idioms such as multiplexer patterns, decoder patterns, and shifter structures. These features highlight higher-level design intent and common microarchitectural motifs.

\noindent \textit{\underline{\textbf{(7) Operator and Expression Features:}}} We count the occurrences of arithmetic, bitwise, and logical operators. These features provide insight into the design computational nature.

\noindent \textit{\underline{\textbf{(8) Lexical and Naming Features:}}} We incorporate lexical style indicators, including identifier frequency, naming conventions, prefix/suffix usage, and ratio of identifiers. Such features reflect coding practices and organizational style. 

\noindent \textit{\underline{\textbf{(9) Graph Features:}}} We extract structural descriptors from graph representations of the design (control flow \& data flow), such as graph size, edge density, and depth. These features capture global structural complexity.

For components (2)–(9), we map the raw features into fixed-length embeddings using a \texttt{FeatureHasher} \cite{featurehasher}, which efficiently handles sparse and categorical data. Each set of features thus becomes a compact embedding vector. The final \textbf{multi-faceted embedding} is obtained by concatenating the semantic embedding from (1) with the hashed feature embeddings from (2)–(9). We quantify the similarity between two designs by evaluating the cosine similarity of their corresponding multi-faceted embedding vectors.

\section{Results and Evaluation}

\subsection{Impact of SAE-Based Interventions on Model Activations}

To evaluate the effect of our SAE-based steering mechanism, we measured the change in model activations across several layers during inference. We hooked into the residual stream of layers 18 through 28 and calculated the L2 norm of the delta between the original and the steered activation vectors. This "delta norm" provides a quantitative measure of the intervention's magnitude at each layer. The experiment was run over multiple inference steps, and the results, summarized in \ref{tab:delta_norm_stats}, demonstrate a consistent and layer-specific impact.
The intervention's magnitude varies significantly across layers. Layer 26 consistently exhibits the largest effect, with a mean delta norm of approximately 56.88. This suggests that the features targeted by our SAE are most prominently represented or influential in this layer. The standard deviation of the delta norms for each layer is remarkably low (typically <= 0.02), indicating that the intervention produces a stable and predictable effect on the model activations across different runs. While Layer 26 is the most affected, the intervention has a substantial impact on a range of layers from 18 to 28. This supports the hypothesis that complex concepts, such as memorized information, are distributed across multiple layers of the model. The lowest impacts were observed in layers 19 and 23. These results provide strong quantitative evidence that our SAE-based steering method can be used to precisely and consistently manipulate model activations in targeted layers.

\begin{table}[b]
\centering
\small
\setlength\tabcolsep{2pt}
\caption{Layer-wise $\Delta$ Norm Statistics}
\label{tab:delta_norm_stats}
\begin{tabular}{@{} l *{8}c @{}}
\toprule
\textbf{Layer} & \textbf{Mean $\Delta$ Norm} & \textbf{Std. Deviat.} & \textbf{Min $\Delta$ Norm} & \textbf{Max $\Delta$ Norm} \\
\cmidrule(r){1-1} \cmidrule(r){2-2} \cmidrule(r){3-3} \cmidrule(r){4-4} \cmidrule(r){5-5}
18 & 27.53 & 0.03 & 27.49 & 27.57 \\
19 & 22.04 & 0.05 & 22.00 & 22.12 \\
20 & 37.59 & 0.04 & 37.54 & 37.64 \\
21 & 28.98 & 0.06 & 28.92 & 29.07 \\
22 & 30.81 & 0.03 & 30.77 & 30.85 \\
23 & 21.64 & 0.01 & 21.62 & 21.65 \\
24 & 34.86 & 0.03 & 34.83 & 34.91 \\
25 & 38.86 & 0.01 & 38.86 & 38.87 \\
26 & \textbf{\underline{56.88}} & 0.10 & 56.77 & 57.00 \\
27 & 32.43 & 0.04 & 32.39 & 32.48 \\
28 & 41.50 & 0.05 & 41.42 & 41.55 \\
\bottomrule
\end{tabular}
\end{table}

\subsection{Identification of Memorization Features}

To identify which SAE features are most associated with memorization of proprietary RTL designs, we conducted a systematic analysis across layers 18-28 of the model. Our approach involved comparing feature activations between proprietary-marked samples and diagnostic (hold-out) samples to isolate memorization-specific patterns.
We identified a total of 275 memorization-critical features distributed across the 11 analyzed layers, representing approximately 0.15\% of all SAE features. As shown in Table \ref{tab:memorization_distribution}, the distribution of these features varies significantly by layer, with layers 26 and 27 containing the highest concentration (33 and 34 features, respectively, or ~0.20-0.21\% of features in those layers). 

\begin{figure*}
\centering
\begin{subfigure}{0.32\textwidth}
\includegraphics[width=\linewidth]{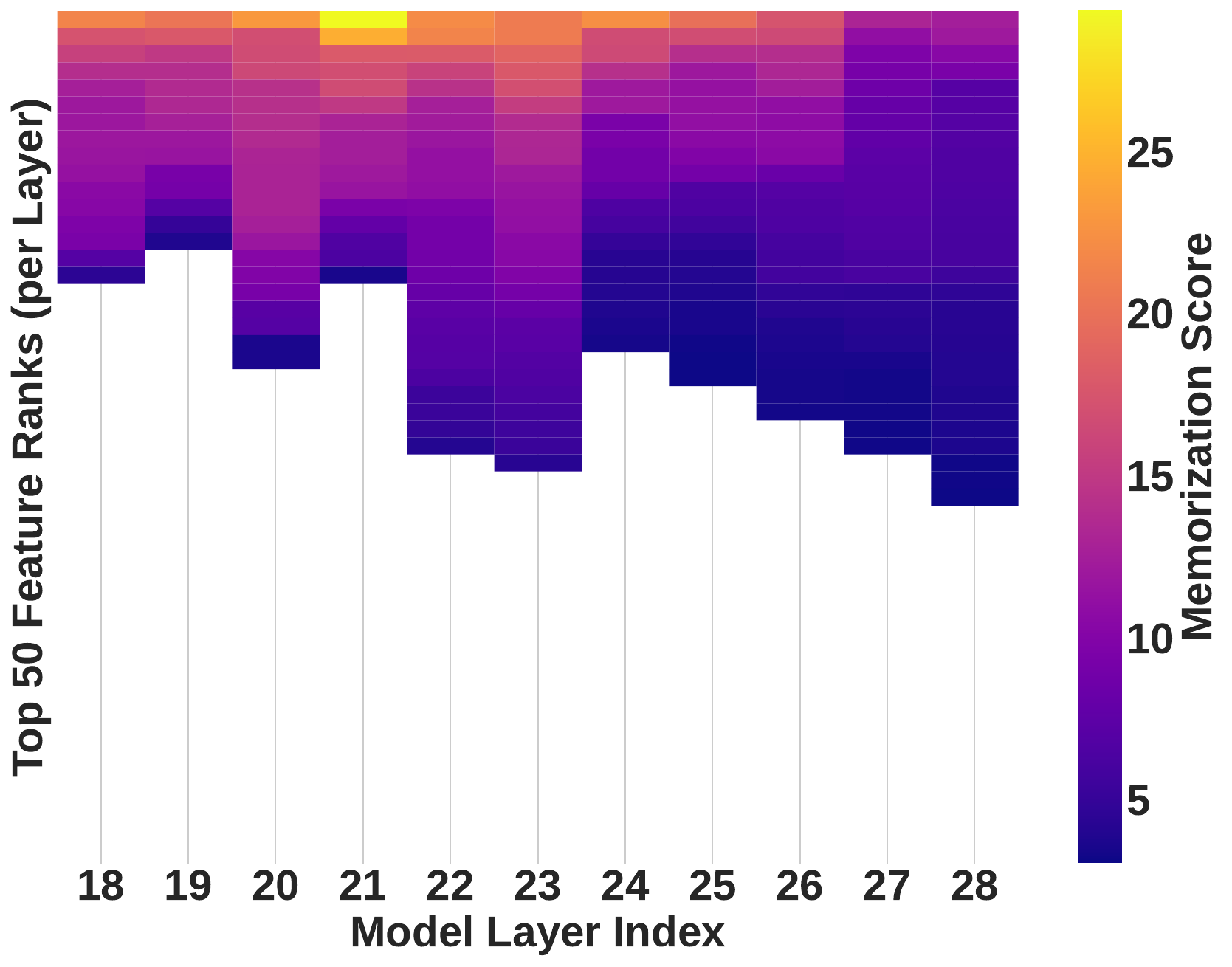}
\caption{Combinational Logic}
\end{subfigure}
\begin{subfigure}{0.32\textwidth}
\includegraphics[width=\linewidth]{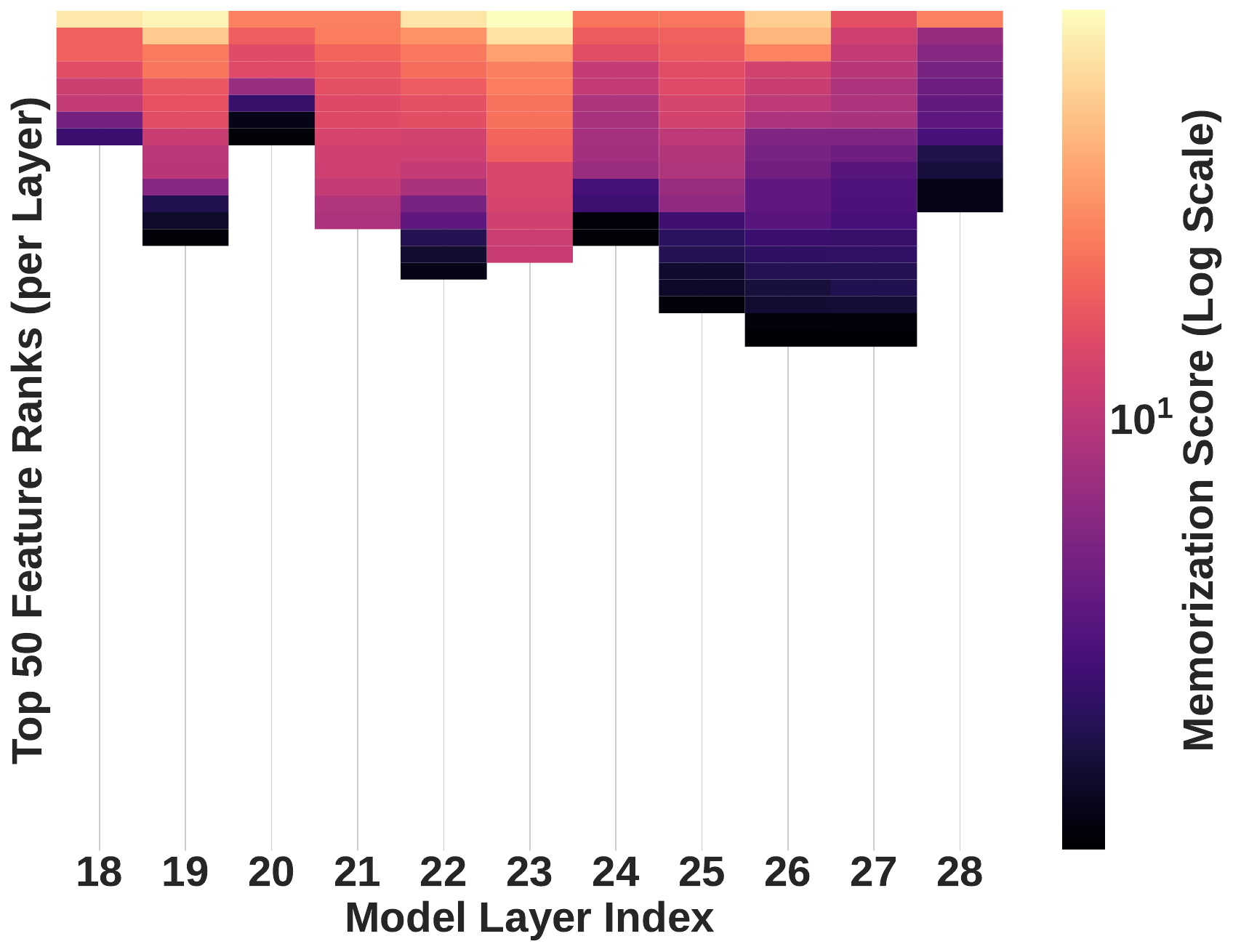}
\caption{Sequential Logic}
\end{subfigure}
\begin{subfigure}{0.32\textwidth}
\includegraphics[width=\linewidth]{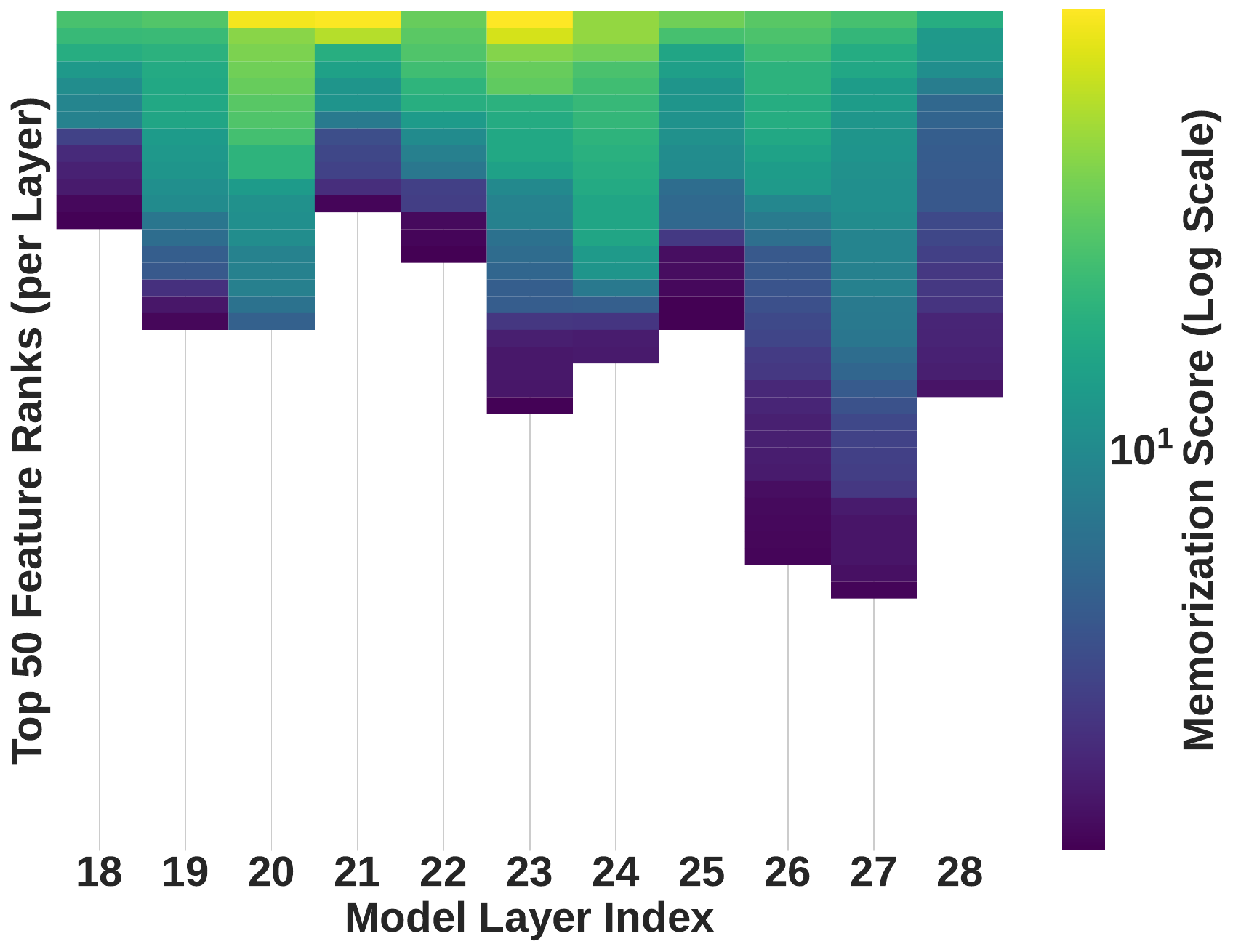}
\caption{Routing/Crossbar}
\end{subfigure}
\caption{Feature Discovery Heatmaps for Categories of RTL Designs.}
\label{fig:feature_discovery}
\end{figure*}

The feature discovery process revealed distinct patterns for different RTL categories. As shown in Figure \ref{fig:feature_discovery}(a), for combinational  logic complex circuits, the concentrated memorization activity is in mid-to-late layers (23-26), with peak scores reaching 25+ on our memorization metric. This suggests that complex combinational designs trigger specific neural pathways primarily in the model's deeper representations. On the other hand, Figure \ref{fig:feature_discovery}(b) illustrates a more distributed pattern with significant activity across layers 20-24, indicating that sequential logic patterns are encoded across multiple abstraction levels rather than being localized to specific layers. For routing/crossbars, as depicted in Figure \ref{fig:feature_discovery}(c), the strongest memorization signals is in layers 25-27, with particularly high concentration in layer 26. The log-scale visualization for routing circuits reveals memorization scores spanning several orders of magnitude, suggesting that multiplexer patterns create highly distinctive activation signatures.

\begin{table}[t]
\centering
\small
\setlength\tabcolsep{4pt} 
\caption{Distribution of Memorization Features Across Layers.}
\label{tab:memorization_distribution}
\begin{tabular}{@{} lcc lcc lcc @{}}
\toprule
\textbf{Layer} & \textbf{Feat.} & \textbf{\%} &
\textbf{Layer} & \textbf{Feat.} & \textbf{\%} &
\textbf{Layer} & \textbf{Feat.} & \textbf{\%} \\
\cmidrule(r){1-3}\cmidrule(r){4-6} \cmidrule(r){7-9} 
\textbf{18} & 17 & 0.10 & \textbf{22} & 19 & 0.12 & \textbf{25} & 30 & 0.18 \\
\textbf{19} & 18 & 0.11 & \textbf{23} & 31 & 0.19 & \textbf{\underline{26}} & \textbf{\underline{33}} & \textbf{\underline{0.20}} \\
\textbf{20} & 32 & 0.20 & \textbf{24} & 22 & 0.13 & \textbf{\underline{27}} & \textbf{\underline{34}} & \textbf{\underline{0.21}} \\
\textbf{21} & 18 & 0.11 & \textbf{28} & 21 & 0.13 \\
\cmidrule(r){1-3}\cmidrule(r){4-6} \cmidrule(r){7-9} 
 & & & & & & \multicolumn{3}{c}{\textbf{Total: 275 (0.15\%)}} \\
\bottomrule
\end{tabular}
\end{table}

\begin{figure}[b]
\centering
\includegraphics[width=\linewidth]{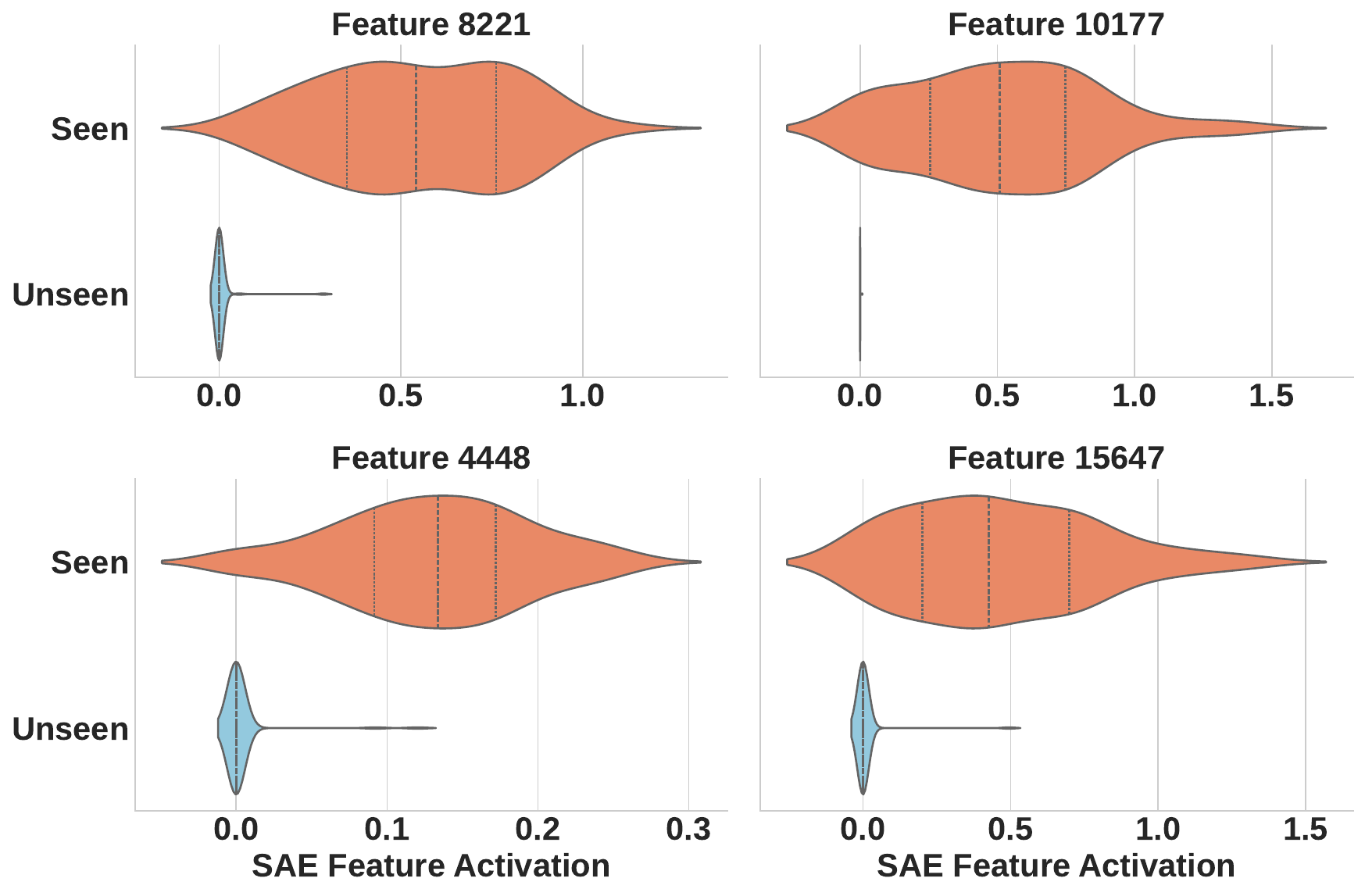}
\caption{Feature Activation Distribution for a Given Mid-Layer (Layer 20).}
\label{fig:feature_activation_distribution}
\end{figure}

Figure \ref{fig:feature_activation_distribution} shows how selected SAE features behave differently on training (seen) vs. held-out (unseen) samples, highlighting the role of memorization in mid-to-late layers (e.g., layer 20). As shown, clear separation emerges across four representative features at given layer. For instance, Feature 8221 shows a mean shift between distributions, while Feature 10177 exhibits the strongest discrimination. While one feature exhibits a bimodal distribution and another shows near-complete separation, the overall trend is consistent: seen data activates strongly, whereas unseen data remains close to zero. These examples illustrate how different memorization features can specialize, either capturing multiple memorized patterns or acting as highly discriminative detectors. At the architectural level, memorization features are not randomly distributed. Roughly 60\% concentrate in layers 23–27, consistent with prior findings that deeper transformer layers encode more specific, memorization-prone representations, while earlier layers (18–22) emphasize generalizable syntactic and structural patterns. Hence, interventions can be targeted at memorization-critical layers, while preserving broad, general-purpose representations in other stages.


\subsection{Steering Effectiveness Results}

\begin{figure}[t]
\centering
\includegraphics[width=0.95\linewidth]{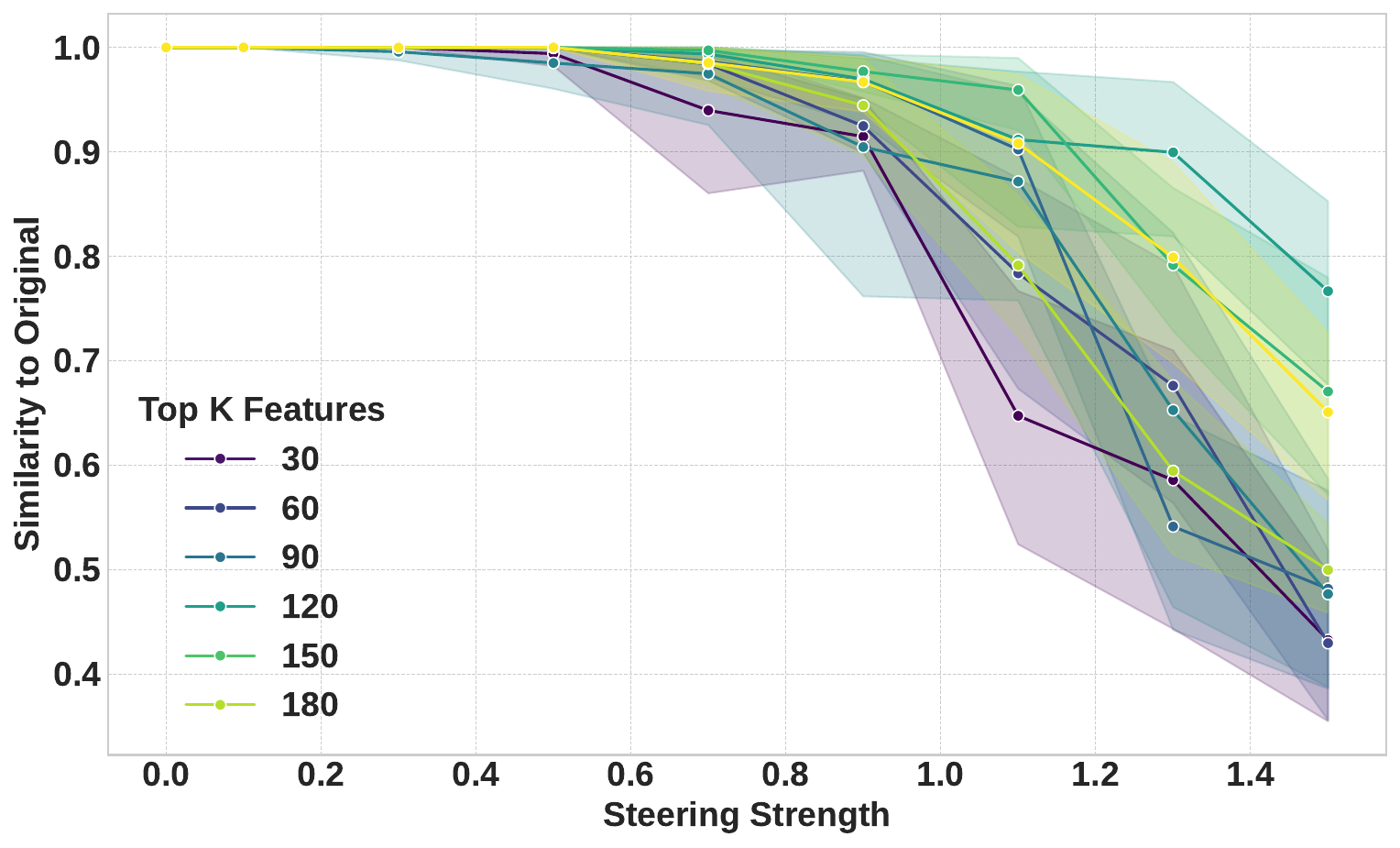}
\caption{The Similarity Ratio vs. Strength of Steering in CircuitGuard.}
\label{fig:similarity_strenght_counter}
\end{figure}

We evaluated CircuitGuard’s steering mechanism across multiple RTL categories by varying suppression strength and the number of features suppressed on randomly selected samples. Figure \ref{fig:similarity_strenght_counter} illustrates results for sequential circuits (counters/shifters), showing how semantic similarity to the original design decreases as steering strength increases. With $K=30$ features suppressed, similarity remains above 0.9 up to strength 0.6, before gradually falling to 0.4 at strength 1.4. Expanding suppression to $K=180$ shifts the curve leftward, indicating stronger memorization reduction at lower steering strengths and demonstrating fine-grained control over the memorization–utility trade-off. Table \ref{tab:cat_sig_points} summarizes critical steering thresholds across categories, highlighting where meaningful memorization reduction first appears (knee point - 10\% semantic difference while maintaining quality scores above 8.0) and where excessive suppression leads to quality loss (oversteer point - 80\% semantic difference with quality degradation below 6.0).

Additionally, CircuitGuard steering effectiveness transfers robustly across RTL categories. For instance, models trained on combinational logic maintained 85\% effectiveness on sequential circuits and 78\% on arithmetic units (see Table \ref{tab:cross_domain_transfer}), with successful transfer rates ranging from 12-15\% across all tested categories. This cross-domain robustness indicates that memorization features capture general patterns rather than category-specific implementations, enabling broad applicability without per-category retraining.

\begin{table}[b]
\centering
\small
\setlength\tabcolsep{2.5pt}
\caption{Significant steering points by category. Knee = first steering with Sem.~$\Delta\!\ge\!10$ and Qual.~$<\!8$; Oversteer = first with Sem.~$\Delta\!\ge\!80$.}
\label{tab:cat_sig_points}
\begin{tabular}{@{} l r l r r r @{}}
\toprule
\textbf{Category} & \textbf{$K$} & \textbf{Point} & \textbf{St} & \textbf{Sem.\,$\Delta$ (\%)} & \textbf{Qual.} \\
\cmidrule(r){1-1} \cmidrule(r){2-2} \cmidrule(r){3-3} \cmidrule(r){4-4} \cmidrule(r){5-5} \cmidrule(r){6-6}
Combinational Logic & 60  & Knee      & 0.9 & 10 & 7.70 \\
                    & 60  & Oversteer & 1.5 & 90 & 4.00 \\
                    & 120 & Knee      & 0.5 & 20 & 7.70 \\
                    & 120 & Oversteer & 1.1 & 80 & 6.00 \\
\cmidrule(r){1-1} \cmidrule(r){2-2} \cmidrule(r){3-3} \cmidrule(r){4-4} \cmidrule(r){5-5} \cmidrule(r){6-6}
Sequential (Counter)             & 20  & Knee      & 0.7 & 40 & 7.40 \\
                    & 20  & Oversteer & 0.9 & 80 & 6.10 \\
                    & 200 & Knee      & 0.9 & 50 & 7.90 \\
                    & 200 & Oversteer & 1.3 & 90 & 5.20 \\
\cmidrule(r){1-1} \cmidrule(r){2-2} \cmidrule(r){3-3} \cmidrule(r){4-4} \cmidrule(r){5-5} \cmidrule(r){6-6}
Routing Modules         & 20  & Knee      & 1.3 & 80 & 7.50 \\
                    & 20  & Oversteer & 1.5 & 90 & 4.60 \\
                    & 180 & Knee      & 0.9 & 50 & 7.50 \\
                    & 180 & Oversteer & 1.1 & 80 & 6.10 \\
\bottomrule
\end{tabular}
\end{table}

\begin{table}[t]
\centering
\small
\setlength\tabcolsep{3pt}
\caption{Cross-domain Transfer (Train: \emph{Combinational Logic}; Test: \emph{Complex}). Columns with all-zero values (Suppressed, Suppression/Retention/Error rates) are omitted. Samples tested per setting: 30.}
\label{tab:cross_domain_transfer}
\begin{tabular}{@{} l *{15}c @{}}
\toprule
\textbf{K >>} & 20 & 40 & 60 & 80 & 100 & 140 & 180 & 200 \\
\cmidrule(r){1-1} \cmidrule(r){2-2} \cmidrule(r){3-3} \cmidrule(r){4-4} \cmidrule(r){5-5} \cmidrule(r){6-6} \cmidrule(r){7-7} \cmidrule(r){8-8} \cmidrule(r){9-9} 
\textbf{Transferred} & 262 & 255 & 275 & 247 & 246 & 255 & 210 & 241 \\
\cmidrule(r){1-1} \cmidrule(r){2-2} \cmidrule(r){3-3} \cmidrule(r){4-4} \cmidrule(r){5-5} \cmidrule(r){6-6} \cmidrule(r){7-7} \cmidrule(r){8-8} \cmidrule(r){9-9} 
\textbf{Transfer Rate (\%)} & 0.15 & 0.14 & 0.15 & 0.14 & 0.14 & 0.14 & 0.12 & 0.13 \\
\bottomrule
\end{tabular}
\end{table}

\subsection{Feature Suppression Analysis}

\begin{figure}[t]
\centering
\includegraphics[width=.95\linewidth]{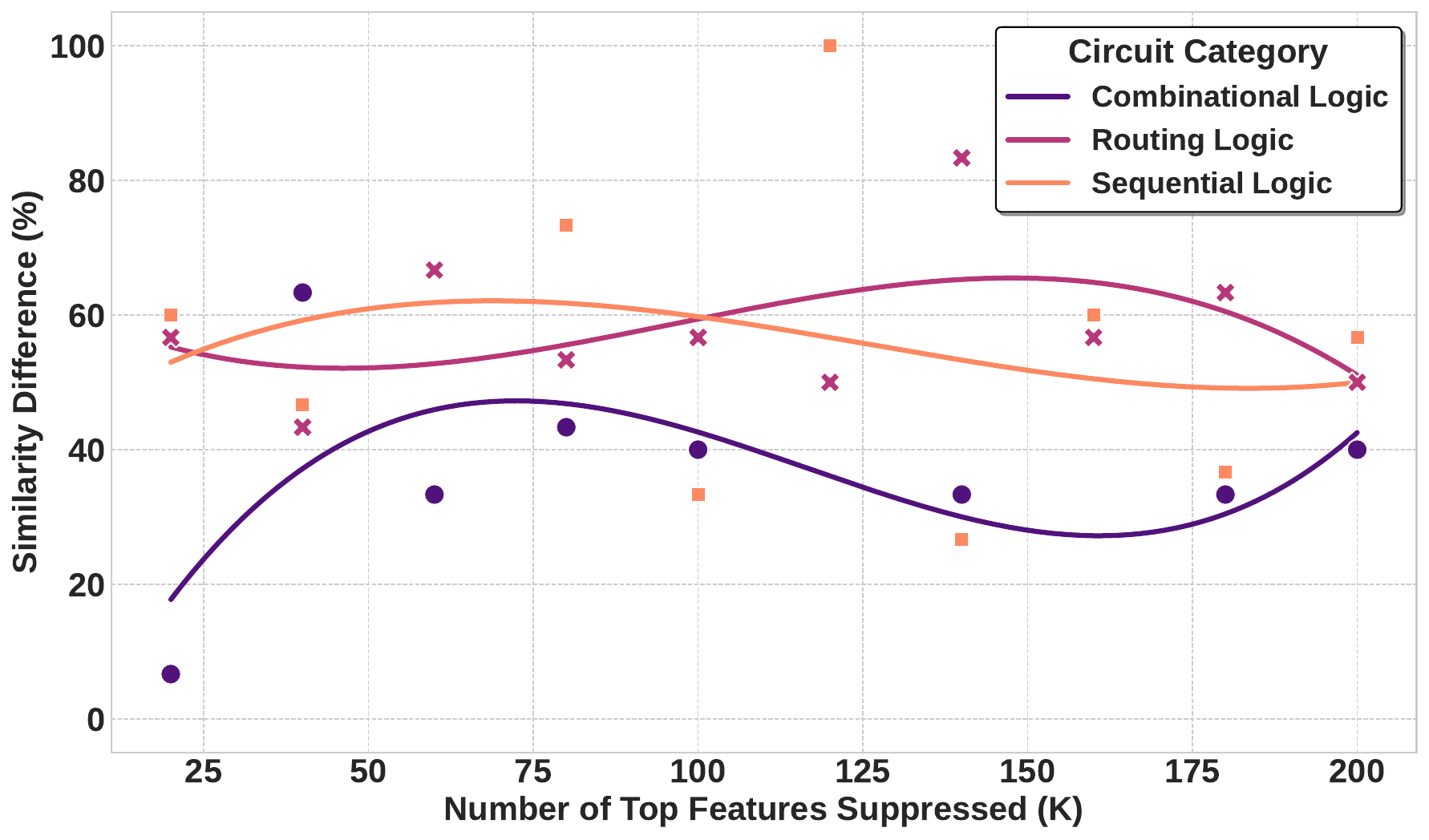}
\caption{Similarity vs Top-K parameter}
\label{fig:k_parameters_sensitivity_analysis}
\end{figure}

Figure \ref{fig:k_parameters_sensitivity_analysis} compares the impact of feature suppression across RTL categories. Sequential circuits emerge as the most sensitive, reaching nearly 80\% semantic difference with suppression of only $K=50$ features. This reflects their reliance on precise state transitions, which are easily disrupted when memorization-linked features are attenuated. Combinational logic circuits show a more gradual response, crossing 50\% difference around $K=100$ features and plateauing near 80\% at higher suppression levels. This suggests memorization is distributed across multiple redundant pathways, requiring broader intervention. Routing blocks prove most robust, with significant differences appearing only beyond $K=120$, consistent with their structural flexibility and multiple valid implementations. Together, these profiles show that sequential designs benefit from aggressive suppression at small $K$, while combinational and routing logic require broader but gentler interventions ($K=100$–150) to balance memorization reduction with functional fidelity.

\subsection{Adaptive Steering}

The main steps of the adaptive steering implementation has shown in Algorithm \ref{alg:adaptive_steering}. CircuitGuard achieves adaptiveness by calculating a memorization risk score for each input and using it to determine an appropriate steering strength before generation. For every prompt, MLP-input activations are captured from layers 18-28 to compute a risk signal that indicates the likelihood of memorization. This risk score is then mapped to an adaptive steering strength.
In practice, CircuitGuard dynamically adjusts steering strength based on the input's characteristics, applying lighter interventions for novel inputs with low memorization risk and stronger interventions for inputs showing higher memorization risk. During inference, the steering mechanism operates through forward hooks at target layers. These hooks implement a decode-difference approach using SAEs: they encode the activations, apply weighted suppression to identified memorization features based on the computed strength, then decode and combine the modified signal with the original activations. Rather than using binary on/off suppression, CircuitGuard scales memorization features by (1 - strength), creating a gradual, context-sensitive intervention. CircuitGuard successfully balances memorization mitigation with functional preservation across 1,552 identified features distributed over 11 layers, demonstrating that input-specific, graduated interventions can reduce memorization risks while maintaining code generation quality.

\begin{algorithm}[t]
\caption{Adaptive SAE Steering (Overall)}
\label{alg:adaptive_steering}
{\small\begin{algorithmic}
\REQUIRE Model $\mathcal{M}$, tokenizer $\mathcal{T}$, SAE set $\{\mathrm{SAE}_\ell\}$, memorization features $\{\mathcal{F}_\ell=(\text{idx}_\ell,\text{score}_\ell)\}$, input tokens $\bm{x}$, generation args $\Gamma$, base strength $s_0$, steps $S$, range $[s_{\min}, s_{\max}]$
\ENSURE Best output $\hat{\bm{y}}$ with quality score $Q(\hat{\bm{y}})$

\STATE \textcolor{red!80!black}{\textbf{// Compute input-adaptive base strength}}
\STATE $s_{\mathrm{adapt}} \gets \textsc{ComputeAdaptiveStrength}(\bm{x}, s_0)$
\STATE $s_{\mathrm{start}} \gets \max(s_{\min}, s_{\mathrm{adapt}})$,\quad $s_{\mathrm{end}} \gets s_{\max}$
\STATE $(\hat{\bm{y}}, \hat{q}) \gets (\varnothing, -\infty)$

\STATE \textcolor{red!80!black}{\textbf{// Progressive sweep of strengths with early stop}}
\FOR{$t \gets 0$ \textbf{to} $S-1$}
    \STATE $s_t \gets s_{\mathrm{start}} + \dfrac{t}{\max(1,S-1)}\big(s_{\mathrm{end}}-s_{\mathrm{start}}\big)$
    \STATE $\bm{y}_t \gets \textsc{ApplySteering}(\mathcal{M}, \{\mathrm{SAE}_\ell\}, \{\mathcal{F}_\ell\}, \bm{x}, \Gamma, s_t)$
    \STATE $q_t \gets \textsc{EvaluateQuality}(\bm{y}_t)$
    \IF{$t>0$ \textbf{and} $q_t < 0.8 \cdot q_{t-1}$}
        \STATE \textbf{break} \COMMENT{Early stop on sharp quality drop}
    \ENDIF
    \IF{$q_t > \hat{q}$}
        \STATE $(\hat{\bm{y}}, \hat{q}) \gets (\bm{y}_t, q_t)$
    \ENDIF
\ENDFOR

\STATE \textbf{return} $(\hat{\bm{y}}, \hat{q})$
\end{algorithmic}}
\end{algorithm}

\subsection{Code Quality and Functionality}

To evaluate whether CircuitGuard preserves the functional correctness and utility of generated RTL code while reducing memorization, we conducted comprehensive code quality assessments using a 1-10 scale across syntactic correctness, functional accuracy, and synthesis compatibility.

\begin{figure}[b]
\centering
\includegraphics[width=.95\linewidth]{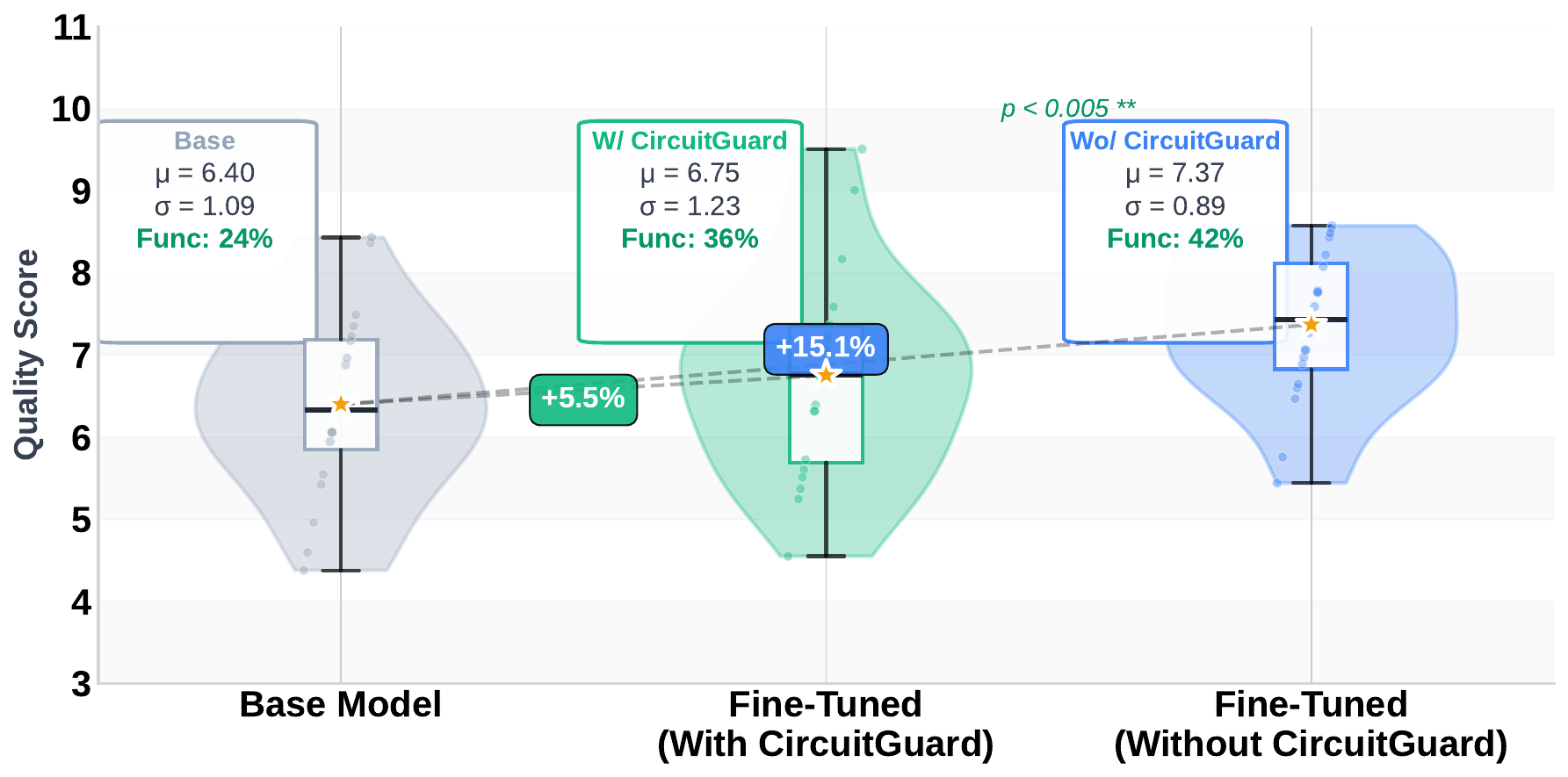}
\caption{Code Quality and Functionality in the Base Model vs. Post-Tuning.}
\label{fig:code_quality_base_vs_ft}
\end{figure}

As shown in Figure \ref{fig:code_quality_base_vs_ft}, the fine-tuned model achieves a mean quality score of 7.37 compared to 6.40 for the base model, representing a +15.1\% improvement in overall code quality. When CircuitGuard steering is applied, the quality score is 6.75, showing an improvement over the base model while successfully mitigating memorization risks. Notably, functional correctness rates show parallel trends: the base model achieves 24\% functional correctness, fine-tuning without CircuitGuard reaches 42\%, and CircuitGuard-steered generation maintains 36\% correctness—demonstrating that memorization reduction comes with a modest but acceptable trade-off in both quality and functionality. CircuitGuard alters the implementation in ways that preserve functionality while reducing memorization of training examples. The fine-tuned model often reproduces patterns closely aligned with training data. 

\begin{figure}[t]
\centering
\includegraphics[width=0.95\linewidth]{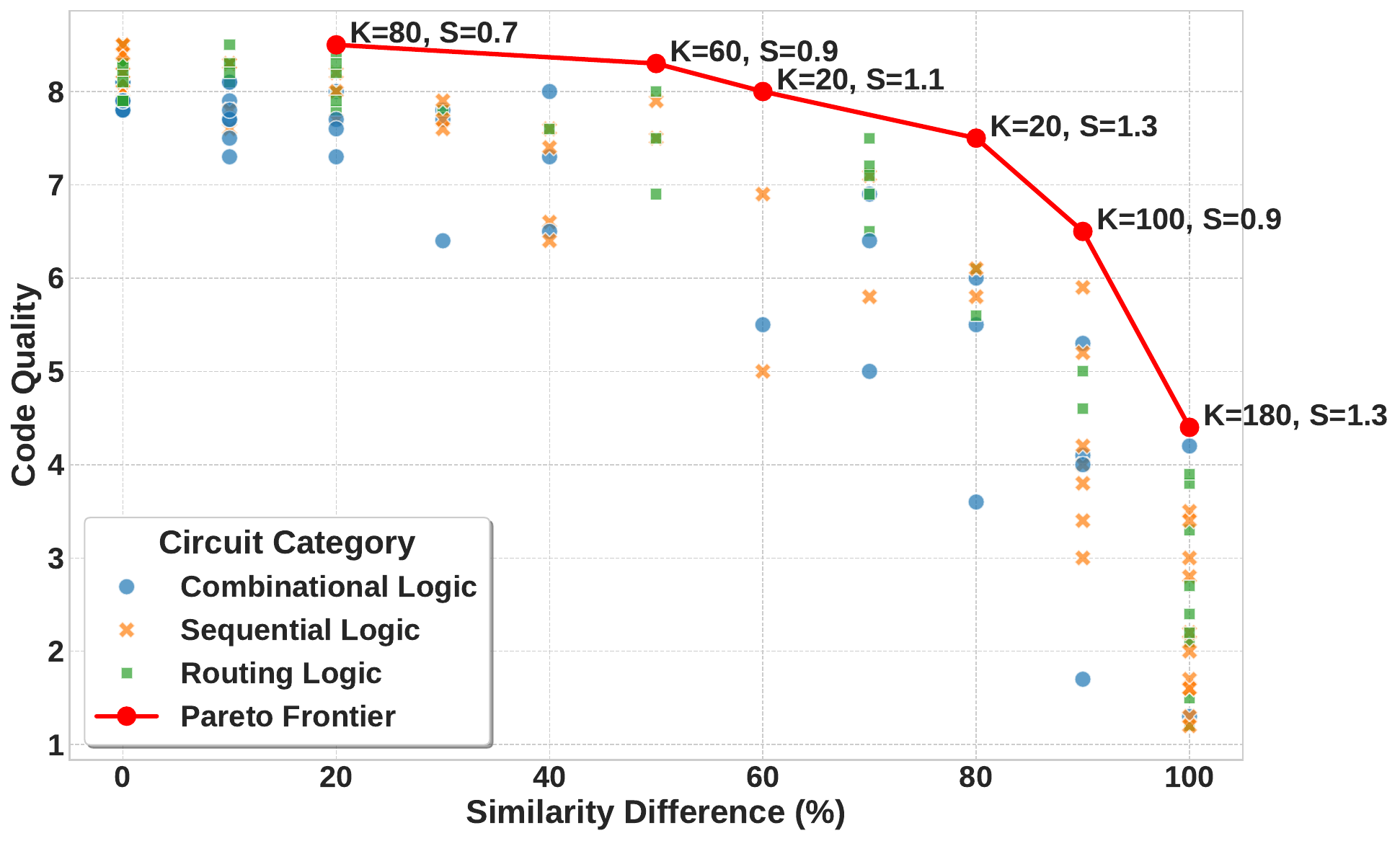}
\caption{Code Quality vs. Memorization Trade-off.}
\label{fig:code_quality_mem_tradeoff}
\end{figure}

We also explored the trade-off between quality and memorization. Figure \ref{fig:code_quality_mem_tradeoff} presents the trade-off between memorization reduction and code quality across different steering configurations. Our results reveal operating points where substantial memorization reduction (40-80\% similarity difference) is achieved while maintaining decent to moderate code quality. For combinational Logic problems, K=60, S=0.9 achieves 10\% memorization reduction with a quality score of 7.7. Different circuit types show varying sensitivity to steering interventions, with sequential circuits demonstrating high memorization reduction. Even under aggressive steering, quality degradation follows a predictable pattern.

\section{Conclusion and Future Works}

In this work, we addressed the critical security challenge of IP leakage in LLM-driven hardware design by introducing CircuitGuard, a defense framework that combines attention-steered unlearning with RTL-aware similarity metrics. CircuitGuard identifies 275 memorization-critical features across transformer layers and achieves up to 80\% reduction in semantic similarity to proprietary patterns. The system demonstrates cross-domain transfer effectiveness of 78-85\% and provides configurable trade-offs between memorization reduction and generation utility. Future research directions include investigating dynamic feature selection, developing more sophisticated RTL similarity metrics with formal verification, exploring integration with differential privacy techniques, and extending the framework to other structured domains.

\bibliographystyle{IEEEtran}
\bibliography{references}

\end{document}